\documentclass[conference]{IEEEtran}

\usepackage{amsmath,amsfonts}
\usepackage{todonotes}
\usepackage{xcolor}

\usepackage{array}
\usepackage{framed}

\makeatletter
\def\endthebibliography{%
	\def\@noitemerr{\@latex@warning{Empty `thebibliography' environment}}%
	\endlist
}
\makeatother

\begin{document}

\title{A Testability Analysis Framework for Non-Functional Properties}

\author{\IEEEauthorblockN{Michael Felderer\\Bogdan Marculescu}
\IEEEauthorblockA{\textit{Blekinge Institute of Technology}\\
Karlskrona, Sweden \\
michael.felderer@bth.se}
\and
\IEEEauthorblockN{Francisco Gomes de Oliveira Neto\\ Robert Feldt \\ Richard Torkar}
\IEEEauthorblockA{\textit{Chalmers and the University of Gothenburg}\\
Gothenburg, Sweden}
}

\maketitle

\begin{abstract}
This paper presents background, the basic steps and an example for a testability analysis framework for non-functional properties.
\end{abstract}


%fon: moved each text to its own file.
\section{Introduction}
%Testing has always been one of the most widely practised techniques for quality assurance of software systems and services in industry. Due to the ever growing complexity and use of software and the, often, limited resources for testing, effective and efficient approaches to testing are highly required. 
Testability is a quality attribute that evaluates the effectiveness and efficiency of testing: If the testability of a software artifact is high, then finding faults by means of testing is easier. A lower degree of testability results in increased test effort, and thus in less testing performed for a fixed amount of time~\cite{voas1995software}. 

%Furthermore, due to the ever increasing integration of software with the social and physical world, quality aspects such as performance, safety, security, and robustness become more important for software systems. Such quality characteristics, which are referred to and captured as non-functional properties and complement the software functionality, determine the success of software products and require special testing approaches.
While software testability has been extensively investigated---in a recent systematic literature review the authors identified 208 papers~\cite{garousi2018testability}---the focus has always been on functional testing, while non-functional properties are often neglected~\cite{hassan2015testability}. Little is known regarding testability of non-functional properties. Thus, there is ample opportunity to investigate the relationship between software testability and different non-functional properties. In this paper we contribute to this unexplored field by characterising and exemplifying a testability analysis framework for non-functional properties. The aim of such an analysis framework is to predict and allocate test resources, assist in the testability design, compare testing approaches or, more generally, to support decision making in projects. The framework is developed based on an in-depth analysis of available testability definitions, testability frameworks and work testability of non-functional properties.

\section{Background and Related Approaches} \label{sec:background}

In this section, we present background on testability definitions, related testability measurement frameworks as well as related work on software testability and non-functional properties. From each part of the section, we draw some conclusions (shown in boxes at the end of each subsection) to guide the development of testability measurement frameworks for non-functional properties.

\subsection{Testability Definitions} \label{sec:testability-def}

Software testability is now established to be a distinct software quality characteristics~\cite{iso25010}. However, testability has always been an elusive, context-sensitive concept and its correct measurement is a difficult exercise~\cite{mouchawrab2005measurement}. Therefore, the notion of software testability has been subject to a number of different interpretations by standards and experts. In their systematic review on software testability, Garousi et al.~\cite{garousi2018testability} provide, overall, 33 definitions for testability extracted from different papers and standards.

A comprehensive testability definition is provided in the ISO\slash IEC Standard 25010 on system and software quality models. It defines testability as the \emph{degree of effectiveness and efficiency with which test criteria can be established for a system, product or component and tests can be performed to determine whether those criteria have been met}. The definition refers to the effectiveness and efficiency aspects of testability and makes explicit that testability is context-dependent with respect to the applied test criteria and the relevant artifacts under test.  

Some testability definitions explicitly cover the efficiency aspect, e.g., when defining testability as \emph{the effort required to test software}~\cite{iso24765}, or the effectiveness aspect, e.g., \emph{measure of how easily software exposes faults when tested}~\cite{yu2016predicting}. 

Other testability definitions define it explicitly via the core testability factors of observability and controllability, e.g., when defining (domain) testability as \emph{ease of modifying a program so that it is observable and controllable}~\cite{poston2012software} .

Finally, there are also testability definitions that provide a more holistic view and also take human and process aspects of testability into account. This is for instance the case in the testability definitions \emph{how easy it is to test by a particular tester and test process, in a given context}~\cite{bach1999heuristics} and \emph{property of both the software and the process and refers to the easiness for applying all the [testing] steps and on the inherent of the software to reveal faults during testing}~\cite{letreon1999self}. 

\begin{framed}
	\begin{itemize}
		\item[TD1] Testability is relative to the test criteria and artifacts under test.
		\item[TD2] Testability is determined by effectiveness and efficiency measures for testing.		
		\item[TD3] Testability has product, process and human aspects.
	\end{itemize}
\end{framed}

\subsection{Available Testability Measurement Frameworks} \label{sec:testability-frameworks}

Most available work on testability provides specific techniques or methods~\cite{garousi2018testability}. But also models, metrics and frameworks are available. In this section, we summarise three relevant and representative empirical frameworks for testability based on the collection provided in~\cite{garousi2018testability} that support testability measurement. 
% We do not consider formal models of testability (e.g.,~\cite{rodriguez2014general}), which are not relevant in our context.

Binder~\cite{binder1994design} provides a testability framework for object-oriented systems. In~\cite{binder1994design} the author claims that testability is a result of six high-level factors: (1) Characteristics of the representation, (2) characteristics of the implementation, (3) built-in test capabilities, (4) the test suite, (5) the test support environment, and (6) the software development process. Each factor is further refined to sub-characteristics, for which occasionally also metrics and relationships are defined. For instance, structure is one sub-characteristic of implementation with assigned complexity metrics like number of methods per class.% For the number of methods per class an upper bound of 20 is suggested.

Mouchawrab et al.~\cite{mouchawrab2005measurement} provide a well-founded measurement framework for object-oriented software testability. The main aim of the framework is to improve testability during software design based on UML diagrams. For each testing phase, i.e., unit, integration, system, and regression testing, attributes that potentially have an impact on software testability in that phase, are provided. For each testability attribute, a list of measurable sub-attributes is defined. For instance, for unit testing the testability attribute unit size with the metrics local features and inherited features (measured for class diagrams) is defined. The framework is complemented by a theory and its associated hypotheses. For instance, one hypothesis states that \emph{increasing the number of local features to be tested increases the cost of unit testing as more test cases are likely to be required and oracles may increase in complexity if they need to account for additional attributes}. 

Bach~\cite{bach1999heuristics} defines five `practical' testability types, i.e., epistemic testability (``How narrow is the gap between what we know and what we need to know about the status of the product under test''), value-related testability (``Testability influenced by changing the quality standard or our knowledge of it''), project-related testability (``Testability influenced by changing the conditions under which we test''), intrinsic testability (``Testability influenced by changing the product itself''), and subjective testability (``Testability influenced by changing the tester or the test process''). For each testability type, characteristics are defined, e.g., domain knowledge or testing skills for subjective testability, and observability and controllability for intrinsic testability. Furthermore, relationships like \emph{improving test strategy might decrease subjective testability or vice versa} are defined.

None of the available testability frameworks examines testability, and its relationship to other non-functional properties, in any details. However, we can draw some conclusions to guide the development of testability analysis frameworks:

\begin{framed}
	\begin{itemize}
		\item[TF1] Testability frameworks define testability characteristics and respective metrics for a specific testability context.
		\item[TF2] Testability frameworks define statements to put the testability context, characteristics and metrics in relations to each other.
	\end{itemize}
\end{framed}

\subsection{Software Testability and Non-Functional Properties} \label{sec:rel-testability-nfp}

As highlighted before, software testability and their relationship to non-functional properties is a relatively unexplored field. However, recently two literature reviews on software testability and its relationship to the non-functional properties robustness~\cite{hassan2015testability} and performance~\cite{hassan2016testability} were published.

The literature review on software testability and robustness includes overall 27 primary studies. The most frequently addressed testability issues investigated in the context of robustness are observability, controllability, automation, and testing effort. The most frequently addressed robustness issues are fault tolerance, handling external influence, and exception handling. Metrics that consider testability and robustness together are rare. In general authors report a positive relationship between software testability and software robustness~\cite{hassan2015testability}.

The literature review on software testability and performance includes overall 26 primary studies. The most frequently addressed testability issues investigated in the context of performance are observability, controllability, automation, and testing effort. Note that the most frequently addressed testability issues in the context of robustness and performance are identical. The most frequently addressed performance issues are timeliness, response time, and memory usage. Again metrics that consider testability and performance together are rare. Furthermore, Gonz{\'a}lez et al.~\cite{gonzalez2009model} present a measurement framework for runtime testability of component-based systems that is related to testability and performance. As runtime testing is, different from traditional testing, performed on the final execution environment it interferes with the system state or resource availability. The framework therefore identifies the test sensitivity characteristics: component state, component interaction, resource limitations and availability, which determine whether testing interferes with the state of the running system or its environment in an unacceptable way as well as the test isolation techniques state separation, interaction separation, resource monitoring and scheduling that provide countermeasures for its test sensitivity. 

\begin{framed}
	\begin{itemize}
		\item[TN1] The most frequently addressed testability issues for robustness and performance are observability, controllability, automation, and testing effort.
		\item[TN2] The most frequently addressed robustness issues are fault tolerance, handling external influence, and exception handling.
		\item[TN3] The most frequently addressed performance issues are timeliness, response time, and memory usage.			
	\end{itemize}
\end{framed}

\section{Towards a Measurement Framework for Non-Functional Testability}
%\section{Measurement Framework Development for Testability of Non-Functional Properties}
\label{sec:testability-nfp}

In this section, we first present goals of a measurement framework for non-functional properties and then sketch our Testability Causation Analysis Framework taking findings from the previous section into account.

\subsection{Overview and Goals}
Our goal is to develop a measurement framework for non-functional properties based on the findings of the previous section. As testability is a relative concept (see TD1) and has different aspects (see TD3), it is not possible to develop a single measurement system that covers all non-functional properties, aspects and contexts. We need a general framework that can be adapted to these points of variation and be instantiated to provide guidance to conceptualise, analyse and measure testability in specific situations.

Available frameworks that have been successfully applied and evaluated for functional testability analysis often take a layered approach and add detail for a specific testability context, set of characteristics and related variables (see TF1). Based on our analysis of existing frameworks above, we adapt and extend the object-oriented testability framework (OOTF) of Mouchawrab et al.~\cite{mouchawrab2005measurement} to address testability of non-functional properties. Their framework is practical and can be used both for approximate, qualitative assessment of testability (`Would testability increase or decrease, given a certain change?'), and as a basis for more exact, quantitative assessment (`How much will testability increase or decrease given a change of this size in this variable?'). A basic assumption it makes is also that the cost to test to a certain level of quality is a natural and hands-on way to conceptualise testability. We thus reuse some aspects of the framework while adapting, extending and generalising it so that it can be applied not only during the analysis and design stages of object-oriented software but for analysis of non-functional properties on any type of software system.

The reusable elements include the different levels and the decomposition of testability into characteristics, sub-characteristics and attributes. That allows the OOTF framework to be adapted towards specific conditions of testability \cite{mouchawrab2005measurement}. However, it is not obvious that non-functional (NF) properties can be captured in this way.

For example, the OOTF framework distinguishes the different levels of testing (unit, integration, system and regression), thus aggregating attributes from lower levels into the higher levels. That distinction between levels of testing is harder to make, or even not needed, when dealing with NF properties, since NF testing not always apply or differ at all levels of testing. Moreover, the OOTF framework does not clearly include factors that account for other aspects of testability such as process, company\slash environment, or the considered testing techniques.

%We aim to strike a balance between having a flexible framework suitable for different non-functional test situations, and proposing attributes and variables that capture testability with respect to different non-functional properties. 
%rf: Not sure what you are trying to say with "balance" here.

To summarise, our contributions relative to the existing framework, are four-fold: $i$) To generalise from OO software to any type of software system, $ii$) to focus on non-functional testability rather than functional, $iii$) to clarify that the same framework can be used both qualitatively and quantitatively, and $iv$) to consider more types of factors of the situation than only design-related factors of the SUT\@. In the following, we further detail our proposed framework called Testability Causation Analysis Framework.

\subsection{TCAF: Testability Causation Analysis Framework}

TCAF is mostly to be used qualitatively but we see a natural extension to also quantitative use. Our adaptation focuses on analysing testability in terms of the inputs that mediate or directly affect it (e.g., SUT, the test technique(s) being used, human and organisational facets, etc.) and their effects on testability outputs (primarily the cost and effectiveness of testing\footnote{Note that we explicitly exclude efficiency here, since it can be defined as effectiveness divided by cost and is thus indirectly being analysed via its sub-components.}).

For testability outputs, we argue that NF properties are typically not atomic and need to be broken down into sub-characteristics or issues. This allows a more detailed analysis. For instance, if we choose Robustness as the NF property, there are the sub-characteristics identified in the literature review of Hassan et al.~\cite{hassan2015testability}, i.e., exception handling, fault tolerance, and handling of external influences. Depending on the specific NF property and level of detail one wants, these might need to be further sub-divided into characteristics. Once this division has been made we have identified a set of NF attributes. For each attribute we then identify specific testability outputs.

%For each of these, one then has to identify specific cost and effectiveness variables. 

An underlying aspect of testability is to measure the time\slash effort\slash cost needed to perform a certain type of testing~\cite{iso24765,mouchawrab2005measurement}, which we will refer simply as \texttt{TestCost}. Therefore, all NF sub-characteristics should be connected to a cost variable. Conversely, in order to capture effectiveness, i.e. the (quality) level to which the testing of the NF (sub-)characteristic has been achieved, we need attribute-specific variables that will often vary depending on a variety of factors. We refer to those attribute-specific variables as the \textit{extent} of testability, or simply \texttt{TestabilityExtent}.

Note that, in some contexts, the extent can be a binary variable where stakeholders do (or do not) have the necessary instruments and dependencies to test the NF attribute, i.e., it is not necessarily continuous. Other scenarios can be a degree of the extent to which testability can be measured (similar to coverage variables). For instance, a situation where a test technique can only be partially applied would mean a reduced extent of the measured testability. A typical example would be when there is a fixed time or cost for conducting a certain type of testing. 

In brief, our framework thus decomposes testability into several levels, beginning with the non-functional property of interest and then further into, potentially, several levels of sub-characteristics to arrive at the NF attributes we consider. Each such NF attribute is then connected to \textit{testability output variables} (TOVs), i.e., \texttt{TestCost} and \texttt{TestabilityExtent}, that capture aspects of the testability \textit{factors} in terms of cost and extent of testing. The main idea of TCAF is then to consider which input factors that would cause a change in these NF TOVs. These input factors are captured in \textit{testability input variables} (TIVs) that are typically of at least three types: those that capture $i$) the surrounding environment, namely the context (e.g., team configuration, processes used, experience with the used test techniques), $ii$) the system under test (e.g., system complexity, number of test interfaces, number of arguments and types of those interfaces), and $iii$) the test techniques considered (e.g., test optimisation and test generation). Given input factors and the output factors they (potentially) affect how one can then proceed to qualitatively analyse the direction and strength of this causation, or to model it statistically and thus being able to predict those effects.

The following steps further detail how to use the TCAF framework:

\begin{enumerate}
    \item Identify Testability Output Variables (TOVs) specific to the non-functional property considered and its different sub-characteristics. These variables will always include the \texttt{TestCost} variable, but can also have \texttt{TestabilityExtent} variables. \textbf{Outputs:} Layered decomposition of the NF property into NF attributes and testability output variables for each of the attributes.

    \item Identify the set of test techniques to be considered or compared in terms of testability. If it is already given that a certain technique can only reach a certain degree of \texttt{TestabilityExtent} for an NF attribute, they need not be further modelled in subsequent steps. If the test techniques imply specific sub-activities in order to be applied refine the \texttt{TestCost} variables from Step 1 to be specific for each sub-activity (designing, executing, reporting the tests, etc.).

    \item Identify system and context attributes that have an impact on TOVs, and define TIVs for them. The test techniques themselves might also have variation points that lead to additional TIVs to include.

    \item Analyse the effect that TIVs have on TOVs. This can either happen Qualitatively or Quantitatively. For the latter, we need quantification of TIV values as well as statistical modelling of the TOVs based on the TIVs. For the former, one needs experience- or research-based reasoning of the level or direction of effect.
\end{enumerate}

We believe that the TCAF framework can help build a causal model of how different attributes\slash variables of the TIV (e.g., context\slash SUT\slash test techniques) determine different aspects of testability in terms of TOVs, i.e., very much how it has been done in other disciplines~\cite{imbens2015causal}. An added benefit is to be able to quantify those variables and, eventually, statistically model the strength of their effect on testability. Given recent progress on actually analysing causality, rather than simply correlating variables with statistical methods, this would now be realistic~\cite{PetJanSch17}. This feature would be highly relevant and useful for estimating\slash predicting the TOVs related to cost, while it may be harder to quantify and then predict TOVs measuring the testability extent.
\section{Example Application: Robustness Testing}
\label{sec:example_application}

\begin{figure}[htb]
	\centering
		\includegraphics[scale=0.4]{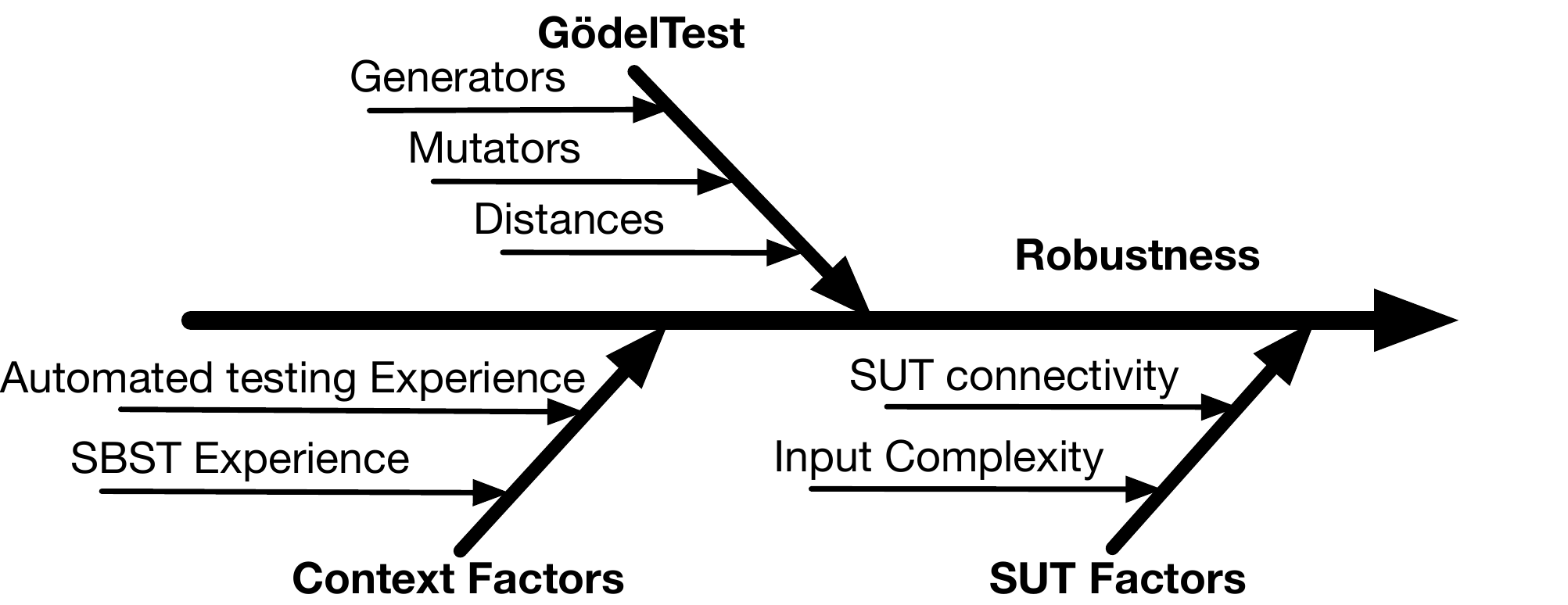}
	\caption{Testability causation analysis example, focusing on robustness.}
	\label{fig:overview}
\end{figure}

This section sketches an example of how TCAF can be applied for testing robustness. An overview of the relevant TIVs is shown in Figure~\ref{fig:overview}.

In the following, we explain each of the four steps to instantiate TCAF.

\textbf{Step 1}: \textit{Testability output variables}. For this example we will focus on \textbf{robustness}. For the sake of brevity, we only consider two exception handling sub-characteristics of robustness, the system's ability to handle atypical and invalid inputs. The TOVs are the cost for and extent to which we can test the two NF attributes: \texttt{CostAtypical}, \texttt{ExtentAtypical}, \texttt{CostInvalid}, \texttt{ExtentInvalid}.

\textbf{Step 2}: \textit{Test technique}. We consider a single test technique in this example: G\"{o}delTest, a search-based, automated test data generation technique. G\"{o}delTest~\cite{feldt2013godel} has been shown to be useful for robustness testing~\cite{poulding2017controllably}. It defines a method of developing valid test inputs of any complexity using a \emph{generator}, to explore invalid test data using one or more \emph{mutation operators}, all driven by a \emph{distance metric} to assess how ``far away'' the generated inputs are from the typical test cases. Each of the three components (generator, mutation operators, distance metric) needs to be in place for the technique to work, so the \texttt{TestCost} associated with each will be assessed separately. When applying this test technique to a large software under test (SUT) we can further consider all these factors for each and every of the interfaces of the SUT that we want to test for robustness, but for the sake of this example we only consider one interface.

\textbf{Step 3}: \textit{System and context variables}. An example of a context attribute that would have an impact on the cost of adopting the technique is that of the relative experience that the company and its testers and developers have with automated testing tools in general, and with search-based software testing (SBST) tools and G\"{o}delTest, in particular. The more experienced the testers and developers are, and the more experienced the company is in developing and using automated testing tools, the lower the costs are likely to be. In addition, the complexity of the SUT is also likely to be an important factor. For example, cost is likely to increase with the number and relative complexity of input data types. For example, it is clear that it is much easier to define a generator for arrays of integers than for graphs represented in XML files.

\textbf{Step 4}: \textit{Causal effects}. The effects can be analyzed depending on the amount of information available, and this analysis can be updated in time. An initial evaluation would most likely be qualitative, focusing on whether each of the TIVs has an effect, and if that effect is likely to be positive or negative. A company may conclude that it does not have many testers or developers with SBST experience, and that is likely to have a negative impact on the cost of adopting G\"{o}delTest. Or it might decide that applying robustness testing on all interfaces is not called for and the testing needs to be more focused. As more information becomes available, the analysis can be more refined, first as a qualitative analysis focusing on discrete steps. For example, when looking at the components of G\"{o}delTest, the company may conclude that it has a number of testing tools that allow the generation of inputs for their SUTs. Thus, \emph{generators} are available for a relatively low cost. On the other extreme, mutation operators would likely be custom, incurring significant cost to develop and validate; in particular, if the input data types are complex and company- or system-specific.

While the analysis for \texttt{CostAtypical} and \texttt{CostInvalid} should be quite similar there is a difference in the number and type of mutation operators needed; the mutation operators for generating atypical inputs are much less complex since we are using the generator as is (atypical inputs are still valid and thus should be captured in the way the generator is defined). Similarly, there are many more invalid data than valid, and thus atypical, so \texttt{ExtentInvalid} will have to be much more constrained and will directly affect \texttt{CostInvalid}. This indicates that more complex analysis or statistical modeling might be needed. It is not always the case that testability outputs can be predicted only from the inputs; outputs might sometimes influence each other.

When possible, the analysis would move more toward quantitative assessments and to include more attributes and factors. For robustness, we could consider other robustness aspects from the literature~\cite{shahrokni2013systematic} as well. A company with experience in working with SBST systems, for example, may be able to estimate the cost of implementing G\"{o}delTest quite accurately, as well as have a clearer understanding of the effect the implementation would have on its products. However, regardless of the level of detail used, TCAF can help structure the testability analysis and make it concrete.
\section{Conclusion}\label{sec:concl}

In this paper we present a testability causation analysis framework for non-functional properties. The framework is developed based on available frameworks and review studies on testability, and prototypically applied to robustness testing.

The framework is used in four steps. First, testability output variables including test cost and testability extend are identified. Second, the set of test techniques to be considered is identified. Third, system and context attributes are identified as testability input variables. Fourth, the effect that testability input variables have on testability output variables are analysed.

So far the framework has not been evaluated. In future, we therefore plan to refine and evaluate the testability causation analysis framework for different non-functional properties including robustness, performance, security and energy consumption (as well as their inter-dependence) in different contexts.

\section*{Acknowledgment}

The paper was partly funded by the Knowledge Foundation (KKS) of Sweden through the project 20130085: Testing of Critical System Characteristics (TOCSYC).

\bibliographystyle{IEEEtran}
\bibliography{references}

\end{document}